# A sparse identification approach for automating choice models' specification

Amir Ghorbani[1], Neema Nassir[2], Patricia Sauri Lavieri[3], Prithvi Bhat Beeramoole[4]

[1,2,3]Transport Engineering Group, The University of Melbourne, Australia

[4] School of Civil & Environment Engineering, Queensland University of Technology, Australia

Email for correspondence: ghorbania@student.unimelb.edu.au

## Abstract

The methodology discussed in this paper aims to enhance choice models' comprehensiveness and explanatory power for forecasting choice outcomes. To achieve these, we have developed a data-driven method that leverages machine learning procedures for identifying the most effective representation of variables in mode choice empirical probability specifications. The methodology will show its significance, particularly in the face of big data and an abundance of variables where it can search through many candidate models. Furthermore, this study will have potential applications in transportation planning and policy-making, which will be achieved by introducing a sparse identification method that looks for the sparsest specification ( parsimonious model ) in the domain of candidate functions. Finally, this paper applies the method to synthetic choice data as a proof of concept. We perform two experiments and show that if the functional form used to generate the synthetic data lies in the domain of base functions, the methodology can recover that. Otherwise, the method will raise a red flag by outputting small coefficients ( near zero ) for base functions.

## 1. Introduction and background

Specifying discrete choice models is a complex and time-constrained process requiring analysts to make sensitive modelling decisions that significantly affect the accuracy of predictions and explanatory power. Linear specifications need to be improved in their ability to capture empirical inflexion points, sudden shifts in preferences and complex variable interactions leading to inaccuracies. Nonlinear specifications can capture more information on causality but require predetermined variables and transformations. Popular methods for capturing nonlinearity include variable transformations, but the selection of variables still largely relies on the analyst. Ultimately, the decisions made during the specification process represent hypotheses that significantly affect the results and interpretation of underlying behaviour captured by the model. Furthermore, the analysis of high dimensional datasets can be laborious and challenging, and it is not feasible to conduct exhaustive hypothesis testing, searching for a solution that addresses all data and modelling aspects (Beeramoole et al., 2023). (Han et al., 2020; Ortelli et al., 2021; Paz et al., 2019; Wu et al., 2019; Zhang et al., 2020) are among notable studies in choice model specification, mainly incorporating machine learning and optimisation techniques. For example, (Ortelli, Hillel, Pereira, et al., 2020)defined a multi-objective combinatorial optimisation problem with two objectives and used a variable neighbourhood metaheuristic to generate solutions mimicking human modellers. (Paz et al., 2019) proposed an optimisation framework for mixed logit models' specification. Recent studies suggest that neural networks outperform certain types of DCMs, such as multinomial logit, regarding predictive power and goodness of fit (Han et al., 2020; Hillel et al., n.d.; Nam & Cho, 2020; Zhang et al., 2020). The literature focuses primarily on enhancing the goodness of fit and predictive measures. However, less attention has been paid to obtaining complex closed-form model specifications which offer a distinctive type of generalisation and descriptive ability. Exhaustively searching and finding the optimal functional form in which explanatory variables should be represented in the utility function is a gap in the choice modelling literature and has been partially addressed by some studies (Beeramoole et al., 2023). When exploring choice models, it is essential to consider a methodological framework that specifies models with a more sophisticated functional form that illuminates the behavioural rules involved in decision-making. For example, conditional and piecewise functions could provide more information on behavioural rules if found to be a proper choice. Despite the value of traditional models, there is still a need to develop an efficient approach that exhausts most, if not all, interactions among variables, variable transformations, and nonlinear functional forms to get closed-form expressions for choice models. The proposed methodological framework has the potential to address these objectives altogether by defining a proper base functions library. Above all, since the only assumption in our methodology is the sparsity, if by testing on multiple datasets we get satisfying performance in terms of predictive power and goodness of fit, we can claim that not only we have obtained a more sophisticated specification but also we have discovered the specification that illuminated the underlying behavioural law, which means we can apply the specification for other applications of relevance. This can provide a more comprehensive understanding of the decision-making process and enhance the predictive power of the models as a ramification. Moreover, the fascinating results of applying sparse identification in recovering the *true* specification in dynamical systems were a great motivation for us (Brunton et al., 2016; Rudy et al., 2017) .By *true* specification in the context of dynamical systems, we mean the ability of the specification algorithm to recover the specification that matches the one which is obtained by applying physics law.

In the upcoming sections, we will elaborate on our proposed method that balances the importance of goodness-of-fit measures and a comprehensive, exhaustive search approach to finding the closed-form expression. We will perform simple experiments to show how our method works. We acknowledge that more experiments should be conducted to showcase the full benefits of our methodology over existing methods in the literature, such as fractional split models(Lavieri et al., 2018; Sivakumar & Bhat, 2002). However, as mentioned earlier, we are primarily offering an exhaustive search algorithm through a large number of candidate specifications. The methodology could be safely used alongside other specification algorithms in the literature to obtain an even more thorough search for a closed-form specification.

**Figure 1: Several possible specifications with the hypothetical explanatory variables m and c**

?
$$U = mc^2$$
?
$$U = mc^{2.5}$$
?
$$U = ln(mc^2)$$

## 2. Method

Many natural signals, such as image and audio, are sparse in some specific domains (e.g., after Fourier transformation). Incorporating the sparsity of a signal enables recovery of the signal with few random observations, as standard practice in compressed sensing(Foucart & Rauhut, 2013). In general, finding the sparse domain of a signal requires extensive experiments, and once obtained, it will provide enough information to recover a signal with a relatively small size of randomly sampled components.(Brunton, Proctor, & Kutz, 2016) built on sparse signal recovery and introduced sparse recovery of governing equations (SINDy) for a dynamical system. The sparsity assumption of a signal translates to a governing equation with the fewest possible terms in it. We can also view this approach as sparse regression that seeks to find the most parsimonious model. Since the original methodology was formulated for dynamical systems, there are several challenges to adapting the approach for choice probabilities. Among and above all is the utility and probabilities' unobserved (latent) nature. In this paper, we will propose an identification method based on sparse identification (Brunton et al., 2016) and try to find the best formula for empirical choice probability based on synthetic data. We note that there are other possible formulates of the problem. Our approach integrates domain knowledge and a data-driven method. The former is achieved by defining the base functions library, which requires expert input and intervention at this stage of our research.

We test our algorithm on synthetic data. The main reason is that since the proper functional form is unavailable, testing the algorithm on real data sets does not provide the opportunity to thoroughly test the identification algorithm's accuracy (Sifringer et al., 2020).

## 3. Formulation of the sparse optimisation

Building on (Brunton, Proctor, & Kutz, 2016), which is a machine learning technique that uses sparse regression to find the best set of equations that describe the evolution of a dynamical system incorporating time series data, we formulate the technique for the empirical probability functional form specification. Given a set of aggregated observations along with explanatory variables related to each observation ( e.g, road characteristics) , we formulate the problem coordinates as:

$$O = F\zeta \qquad (1)$$

Where:

$$O = \begin{pmatrix} \widehat{P}_{1n} \\ P_{11} \\ P_{12} \\ . \\ . \\ P_{1J} \end{pmatrix}, F = \begin{pmatrix} f_1=x_1 & f_2=x_2 & f_3=x_3 & f_4=f_1f_2 & . & . \\ x_{11} & x_{12} & x_{13} & . & . \\ x_{21} & x_{22} & x_{23} & . & . \\ . & . & . \\ . & . & . \\ x_{J1} & x_{J2} & x_{J3} \end{pmatrix}, \zeta = \begin{pmatrix} \zeta \\ \zeta_{11} \\ \zeta_{21} \\ ... \\ \zeta_{k1} \end{pmatrix}$$

The optimisation problem to get the sparsest set of governing equations for empirical probability, ***P***, is:

$$Min_x |\zeta|_1 \qquad (2)$$
$$Subject\ to: |F\zeta - O| < \pi$$

With:

**F** = Library (of base functions) matrix

$\zeta$ = Coefficient matrix

**O** = Observed empirical probabilities

$\pi$ = Acceptance threshold

$x_i = i^{th}$ variable such as travel time

J = Number of observations (after aggregation)

k = Number of base functions in the library

$f_n = n^{th}$ base function such as $\ln(x_i)$

The algorithm's output ($\zeta$ matrix) returns the coefficients of the base functions, and once the coefficients are determined, the governing equations can be expressed as:

$$P = \Sigma_j \zeta_j f_j(x) \qquad (3)$$

The objective of the optimisation problem is to find the sparsest possible coefficient vector (parsimonious model) by minimising the L1-norm subject to the exposures (AX) being close enough to the observed values within a certain threshold ($\pi$)(Foucart & Rauhut, 2013). Note that $\pi$ is a regularisation parameter of the problem and determines the trade-off between a parsimonious model and fitting to observation. Also, note that the defined base functions in Eq.[1] are for illustration purposes, and the researcher has free rein to define the library of the base functions. Since we usually work with large datasets, the researcher can input much more entries to the base function and perform a more exhaustive search. In the next section, we will perform some simple experiments so the reader can primarily verify the methodology's results. Also, by adding columns to $P$ and $\zeta$ matrices, we can solve a one-off optimisation problem for the probability of each alternative. Therefore :

$$P_i = \Sigma_j \zeta_{ji} f_j(x) \tag{4}$$

Where $i$ denotes the alternative. The algorithm searches for the specification in the specification space of size $O(2^k)$. The optimisation technique is convex optimisation using CVXPY library in Python. This specification algorithm searches through its unique and vast specification domain and could be used along with other specification methodologies in the literature to obtain the best possible expression.

## 4. Experiments

### 4.1 Synthetic data: Binary mode choice

First, we generate data with a simple utility for a binary mode choice scenario, defined as:

$$V_1 = 2 \cdot x_{[1]} + 3 \cdot x_{[2]} + 0.5 \cdot x_{[3]} \cdot x_{[4]} + x_{[3]} \cdot x_{[5]} \tag{5}$$
$$P_1 = \frac{1}{1 + e^{V_1}}$$

Where $x_{[1,2,3,4,5]} \sim uniform(-1, 1)$ and the choice outcomes are generated using $Choice \sim Bern(P_1)$. Data is then aggregated and inputted into the algorithm. Finally, the algorithm is run based on the library of base functions defined as:

$$f_1 = x_1 \tag{6}$$
$$f_2 = x_2$$
$$f_3 = x_3$$
$$f_4 = x_4$$
$$f_5 = x_5$$
$$f_6 = x_2 x_3$$
$$f_7 = x_3 x_4$$
$$f_8 = x_3 x_5$$
$$f_9 = 2x_1 + 3x_2 + 0.5 x_3 x_4 + x_3 x_5$$
$$f_{10} = \frac{1}{1 + e^{2x_1 + 3x_2 + 0.5 x_3 x_4 + x_3 x_5}}$$

And the optimization method of (Agrawal et al., 2018; Diamond & Boyd, 2016) with $\pi = 0.001$. The choice of the library function is entirely arbitrary, and the $10^{th}$ base function was *deliberately* set to the true ( generative) functional form to test the algorithm on whether it can recover the true functional form or not. The algorithm was run ten times to obtain ten sets of estimated coefficients. Results are depicted in Table[1].

Table 1. Estimation results for base functions coefficients

| Base Functions | Mean-Coef. | t-value | P-value |
|---|---|---|---|
| f1 | 0.001793352 | 0.02737 | 0.978762 |
| f2 | 0.05266041 | 0.729324 | 0.48435 |
| f3 | -0.044906069 | -0.2441 | 0.81263 |
| f4 | 0.032155775 | 0.168547 | 0.86988 |
| f5 | 0.145794998 | 0.831887 | 0.426993 |
| f6 | 0.088872261 | 0.46599 | 0.652295 |
| f7 | 0.009376727 | 0.036508 | 0.971675 |
| f8 | 0.027603344 | 0.101207 | 0.921605 |
| f9 | 0.193859639 | 2.284382 | 0.048216 |
| f10 | 1.053259665 | 8.482461 | 1.38E-05 |

The analysis of the results indicates that the coefficient corresponding to f10 has the highest t-value (8.48) and the lowest P-value (1.382E-05), which suggests that it is the most significant term in the base functions library. The second significant coefficient is related to f9, which represents the power of the exponential function in f10. This finding is consistent with the idea that the algorithm can detect the relevant terms even if they are not in their exact functional form.

These results prove that the algorithm has successfully recovered the actual probability function with reasonable accuracy. By identifying the key factors that influence binary choice outcomes, this approach can help improve decision-making processes in various domains. It is worth noting that the primary output of the algorithm was the sum of coefficients times their base functions, and the P-value enabled us to further reduce the complexity of the specification by omitting non-significant terms.

### 4.2 Synthetic data : Complex choice scenario

In this experiment, we first generate synthetic data with:

$$x_{[0],[2],\ldots,[39]} \sim Uniform(0,100)$$

$$V_1 = 0.5 x_{[31]}^2 + \sqrt{|x_{[33]}|} + x_{[35]}^3 + 2\log(x_{[37]}) + 4 e^{x_{[39]}}$$

(7)

$$V_2 = 0.8 \cdot \sin(x_{[21]}) + 0.6 \cdot \cos(x_{[22]}) + 0.4 \cdot \tanh(x_{[23]}) + 0.2 \cdot e^{x_{[24]}}$$

$$P_1 = \frac{V_1}{V_1 + V_2} \ , \ P_2 = \frac{V_2}{V_1 + V_2}$$

The choice outcomes are generated using random draws from $P_1$ and $P_2$. Data is then aggregated and inputted into the algorithm to specify the empirical probability function. We defined the library of base functions as in Figure [2].

**Figure 2. Base functions acting on explanatory variables ( x1 to x40 ) and their transformations**

```
base_funcs_list = [

{
#
        'yi(x)': [i for i in range(r)], # identity function
       'power(x, 2)': [i for i in range(r)], # x^2
       'power(x, 3)': [i for i in range(r)], # x^3
       'power(x, 4)': [i for i in range(r)], # x^4
       'power(x, 5)': [i for i in range(r)], # x^5
       'sqrt(abs(x))': [i for i in range(r)], # square root function
       # inverse sine function with clipping and shift:
       'np.arcsin(np.clip(x,-1,1))/(np.clip(x,-1,1)+1e-9)': [i for i in range(r)],
       # inverse cosine function with clipping, shift, and avoidance of division by zero :
       'np.arccos(np.clip(x,-1,1))/(np.clip(x,-1,1)+1e-9)': [i for i in range(r) if i != r//2],

       'np.arctan(x)+1e-9': [i for i in range(r)], # inverse tangent function with shift
       # hyperbolic cosine function with clipping and shift:
       'np.cosh(np.clip(x,-700,700))-1': [i for i in range(r)],
       'np.tanh(x)': [i for i in range(r)], # hyperbolic tangent function
       # logarithmic function with clipping, shift, and avoidance:
       'np.log(np.clip(x,1e-9,np.inf))/(np.clip(x,1e-9,np.inf)+1e-9)': [i for i in range(r*8)],
}
]
```

Where $r = 40$( number of variables ), and the list in front of each base function denotes the columns where each base function is applied, the first 40 columns of the library are initiated with $x_0$ to $x_{39}$. For example, the last logarithmic function is applied on columns with index 0 to 320, which means all the columns that have previously been passed through the first eight base functions are now passed through a logarithmic function and added horizontally to the library of base functions matrix.

Moreover, the optimization method is the same as in the previous section with $\pi = 0.01$. A sample of resulting coefficients is depicted in Table[2] after running the algorithm for a single time. We observed that all of the resulting coefficients were close to zero, which means none of the base functions is active in both probability specifications. This makes sense as the true probability specification is a complex fractional form. The findings indicate a requirement for a more exhaustive methodology to include complex fractional functional forms. Although our methodology searches through a vast number of specifications ( $O(2^{759})$ ), it does not find the true(generative) specification within its hypothesis domain. This is because the algorithm does not search through complex fractional functional forms. Addressing this gap will be an excellent future direction for research.

**Table 2. Estimation results for base functions coefficients : most of the coefficients are near zero**

| Base function | P1- coef. | P2- coef. |
|---|---|---|
| f0 | -8.87572E-36 | -1.33549E-35 |
| f1 | -1.49732E-36 | -3.35131E-36 |
| f2 | -1.71364E-36 | -3.1252E-36 |
| f3 | -1.97421E-36 | -6.75385E-36 |
| . | . | . |
| . | . | . |
| . | . | . |
| f756 | -5.62348E-30 | 6.00547E-29 |
| f757 | -5.62348E-30 | 6.00547E-29 |
| f758 | -7.46936E-39 | -3.40389E-38 |

## 5. Conclusion and future direction

In conclusion, this paper has presented an unexplored method for improving choice models' comprehensiveness and explanatory power, performing experiments with synthetic data. By leveraging machine learning processes to specify the most effective representation of variables, this approach will improve our ability to analyse and predict transportation behaviour, particularly concerning big data and a vast array of variables.

Future research directions should include refining and validating the proposed methodology using more complex real-world datasets to evaluate its usefulness and applicability in various transportation contexts (Ghorbani, 2022). Additionally, exploring the integration of our approach with other emerging techniques, such as deep learning and reinforcement learning (Yazdani et al., 2023), could further enhance the modelling capabilities and uncover previously unobserved patterns in choice behaviour. Moreover, extending the methodology to include more complex relations between base functions is a critical research direction. Finally, it would also be valuable to investigate the potential for extending our methodology to other domains beyond transportation planning, such as urban planning and finance, where choice behaviour plays a critical role.

By advancing choice modelling practices, we hope to provide policymakers and planners with more robust tools to handle complex transportation challenges and optimise infrastructure investments.